# Spin Coherence and Spin Relaxation in Hybrid Organic-Inorganic Lead and Mixed Lead-Tin Perovskites

Haochen Zhang[1,2], Zehua Zhai[1], Zhixuan Bi[1], Han Gao[3], Meng Ye[4], Yong Xu[1,5,6], Hairen Tan[3,7], Luyi Yang[1,2,5,6,*]

[1]State Key Laboratory of Low Dimensional Quantum Physics, Department of Physics, Tsinghua University, Beijing 100084, China

[2]Department of Physics, University of Toronto, Toronto, Ontario M5S 1A7, Canada

[3]National Laboratory of Solid State Microstructures, Collaborative Innovation Center of Advanced Microstructures, Jiangsu Key Laboratory of Artificial Functional Materials, College of Engineering and Applied Sciences, Nanjing University, Nanjing 210093, China

[4]Graduate School of China Academy of Engineering Physics, Beijing 100193, China

[5]Frontier Science Center for Quantum Information, Beijing 100084, China

[6]Collaborative Innovation Center of Quantum Matter, Beijing 100084, China

[7]Frontiers Science Center for Critical Earth Material Cycling, Nanjing University, Nanjing 210093, China

[*]e-mail: luyi-yang@mail.tsinghua.edu.cn

## Abstract

Metal halide perovskites make up a promising class of materials for semiconductor spintronics. Here we report a systematic investigation of coherent spin precession, spin dephasing and spin relaxation of electrons and holes in two hybrid organic-inorganic perovskites $MA_{0.3}FA_{0.7}PbI_3$ and $MA_{0.3}FA_{0.7}Pb_{0.5}Sn_{0.5}I_3$ using time-resolved Faraday rotation spectroscopy. With applied in-plane magnetic fields, we observe robust Larmor spin precession of electrons and holes that persists for hundreds of picoseconds. The spin dephasing and relaxation processes are likely to be sensitive to the defect levels. Temperature-dependent measurements give further insights into the spin relaxation channels. The extracted electron Landé g-factors (3.75 and 4.36) are the biggest among the reported values in inorganic or hybrid perovskites. Both the electron and hole g-factors shift dramatically with temperature, which we propose to originate from thermal lattice vibration effects on the band structure. These results lay the foundation for further design and use of lead- and tin-based perovskites for spintronic applications.

TOC Graphic

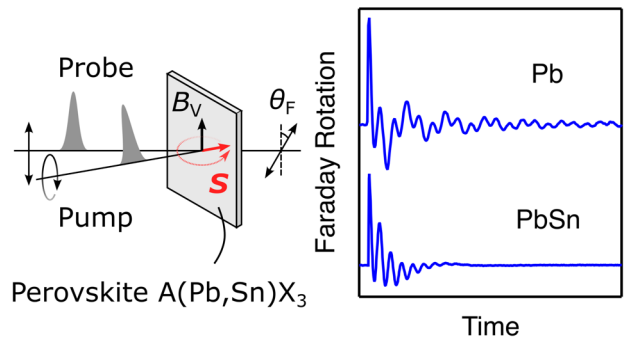

Metal halide perovskites have been intensively studied due to their excellent optoelectronic performance[1], but have only started to gain attention as promising materials for spintronic applications[2]. On top of their unique properties such as great defect tolerance and high absorption coefficient, the efficient spin injection has been demonstrated through optical pumping methods[3–5], and long spin coherence times from hundreds of picoseconds to over one nanosecond have been observed at cryogenic temperatures [4–8]. Furthermore, the Landé g-factor, an important parameter for spin manipulation via external fields, has been investigated in detail in various inorganic and hybrid perovskite systems through ultrafast optical methods[4–9], and their values can be accurately modeled based on the perovskite's band structure parameters, especially their bandgap and spin-orbit coupling (SOC) splitting energy[9–11].

However, most research of spin physics in perovskite semiconductors has been focusing on conventional lead-based systems[3–9,11–14]. Their tin-based counterparts, which were synthesized recently with reasonably high quality[15] and showed top-tier photovoltaic performance in tandem solar cells[1,15], also provide a fertile playground to explore spin properties. First, the tin perovskites have a smaller bandgap and SOC gap than the lead perovskites[16], and can thus test previous models on the perovskite band structure such as the $\boldsymbol{k} \cdot \boldsymbol{p}$ model on the Landé g-factors[10,11]. Second, the smaller SOC might reduce the spin relaxation, leading to a longer spin coherence time, which could be potentially exploited for spin transport[2]. Moreover, the tin perovskites generally have more vacancies and higher trap densities than the lead-based ones[15,16] due to the easy oxidation of tin cations from $Sn^{2+}$ to $Sn^{4+}$, so defect physics is likely important in evaluating their spin properties.

In addition, thermal effects on the spin properties also need to be investigated to put forward a perovskite-based spintronic device at room temperature, and a complete understanding of the thermal evolution of the spin states and the spin relaxation in perovskites is still lacking[3–7,12–14,17–19]. For instance, a strong temperature dependence of the Landé g-factors in a pure-lead perovskite has been observed[4]. Nevertheless, the origin of such an effect has not been analyzed in detail.

In this work, we do a comparative study of the spin coherence and spin relaxation in two hybrid organic-inorganic perovskite thin films $MA_{0.3}FA_{0.7}PbI_3$ and $MA_{0.3}FA_{0.7}Pb_{0.5}Sn_{0.5}I_3$ (abbreviated to Pb- and PbSn-perovskite, respectively, hereafter), where MA = $CH_3NH_3$ and FA = $(NH_2)_2CH$. Using time-resolved Faraday rotation spectroscopy, we detect long-lived spin relaxation and spin coherence of electrons and holes, from which we extract spin lifetimes, Landé g-factors and the inhomogeneous broadening of the g-factors. The electron g-factors are 3.75 and 4.36 in the Pb- and PbSn-perovskites, which are the biggest among the reported g-factors in inorganic or hybrid perovskites. We observe contrasting spin lifetimes between the two samples, suggesting that the spin relaxation is likely due to scattering with defects via the Elliot-Yafet mechanism at low temperatures and the spin decoherence suffers from g-factor inhomogeneity due to impurities and vacancies. By measuring carrier spin lifetimes at elevated temperatures, we specify possible roles of defects and phonons in the spin relaxation channels. Temperature-dependent experiments show drastic changes of both electron and hole g-factors. Supported by a model, we propose, for the first time, that this effect is dominated by the enhancement of dynamic lattice distortions (lattice vibrations) with increasing temperature, resulting in strong modifications of not only the bandgap but also the interband transition matrix and the SOC gap. Our results provide insights for the development of future hybrid perovskite spintronic materials.

The Pb- and PbSn-perovskite samples are solution-processed polycrystalline thin films. Both compounds have a cubic structure and show no structural phase transition from 90 K to room temperature from X-ray diffraction measurements[20], as expected from the optimized ratio of the organic cations MA and FA in our samples[21,22].

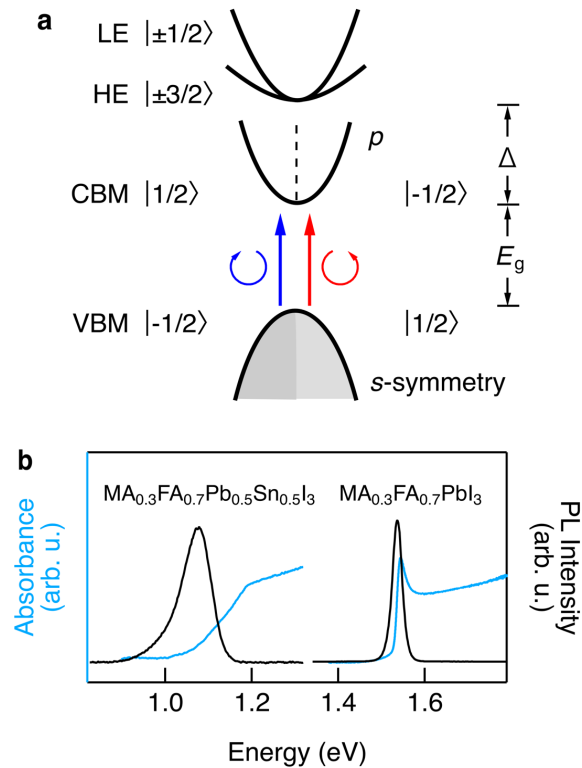

**Figure 1 a** Band structure and spin optical selection rules. The conduction band minimum (CBM) is a spin-orbit coupling split-off band of the Pb/Sn *p*-orbitals with total angular momentum *j* = 1/2, and the light and heavy electron states (LE and HE) are higher energy bands with total angular momentum *j* = 3/2. The valence band maximum (VBM) is formed by the Pb/Sn *s*-orbitals and the I *p*-orbitals retaining *s* symmetry (*j* = 1/2). Therefore, the band edges of the perovskites are formed by effective spin 1/2 states, allowing selective pumping of spin polarized electrons and holes in the system through the absorption of circularly polarized photons. Blue arrow: right circularly polarized light; red arrow: left circularly polarized light. The spin-orbit splitting energy is denoted by *Δ* in the conduction bands and the bandgap is denoted by $E_g$. **b** Absorbance and photoluminescence (PL) spectra of the two samples at 10 K, where arb. u. stands for arbitrary units.

The cubic-phase perovskites are direct gap semiconductors where the gap is located at the R point of the Brillouin zone[10,23]. Figure 1a shows a schematic of the band structure and the spin optical selection rules near the band edge. The conduction band minimum (CBM) is formed by the Pb/Sn *p*-orbitals, and the valence band maximum (VBM) is formed by the Pb/Sn *s*-orbitals and the I *p*-orbitals retaining the *s* symmetry[24]. The SOC causes an energy splitting Δ in the conduction band and the CBM is an SOC split-off band from the higher energy heavy electron (HE) and light electron (LE) bands[23].

Figure 1b shows the absorption and photoluminescence (PL) spectra for the Pb- and PbSn-perovskites at 10 K, from which the band gaps are determined to be 1.56 and 1.18 eV, respectively. Note the sharp absorption edge, pronounced exciton absorption peak and narrow PL width for the Pb-perovskite, in stark contrast to the results of the PbSn-perovskite. This is due to the presence of $Sn^{4+}$ defects and high trap densities in the PbSn sample[15], which also influence the spin lifetime as discussed later.

Efficient spin polarization in the perovskites can be achieved by the optical orientation effect, which is based on optical selection rules subject to the band structure of the perovskites[4,10,23,25]. The CBM and VBM of perovskites consist of effective spin 1/2 states, and therefore circularly polarized light can be used to selectively excite spin-polarized electrons and holes due to the conservation of spin angular momentum, similar to the optical spin selection rules in GaAs[25].

The TOC figure shows a schematic of the time-resolved Faraday rotation experiment. The spin optical selection rules enable the injection of spin-polarized electrons and holes via a circularly polarized pump pulse. The subsequent time evolution is monitored by measuring the Faraday rotation of a time-delayed linearly polarized probe pulse. Both the pump and probe beams are at near normal incidence; therefore, the out-of-plane spin component is initialized and detected. An external magnetic field $B_V$ is applied in the transverse direction (i.e., perpendicular to the laser beams, the Voigt geometry), inducing coherent spin precession along it.

The time-resolved Faraday rotation signals for the Pb- and PbSn-perovskites at 4 K are shown in Figure 2a and b. At zero field, the signal lasts over hundreds of picoseconds for both perovskites. With an applied magnetic field $B_V$, the signal becomes oscillatory indicating spin precession along $B_V$. Both the frequency and the decay rate of the precession signals become bigger as the field is stronger. The Fourier transform of the time traces reveals two oscillation frequencies, both of which increase linearly with the magnetic field strength, as shown in Figure 2c and d. The time-domain raw data can be fit well with two exponentially decaying cosines: $\theta_F(t) = \sum_{i=1}^{2} A_i \cos(2\pi f_i t + \phi_i) \exp(-t/T^*_{2,i})$. We plot the extracted oscillation frequency $f_{e,h}$ and the spin decay rate $1/T^*_{2,e,h}$ in Figure 2e and f, respectively.

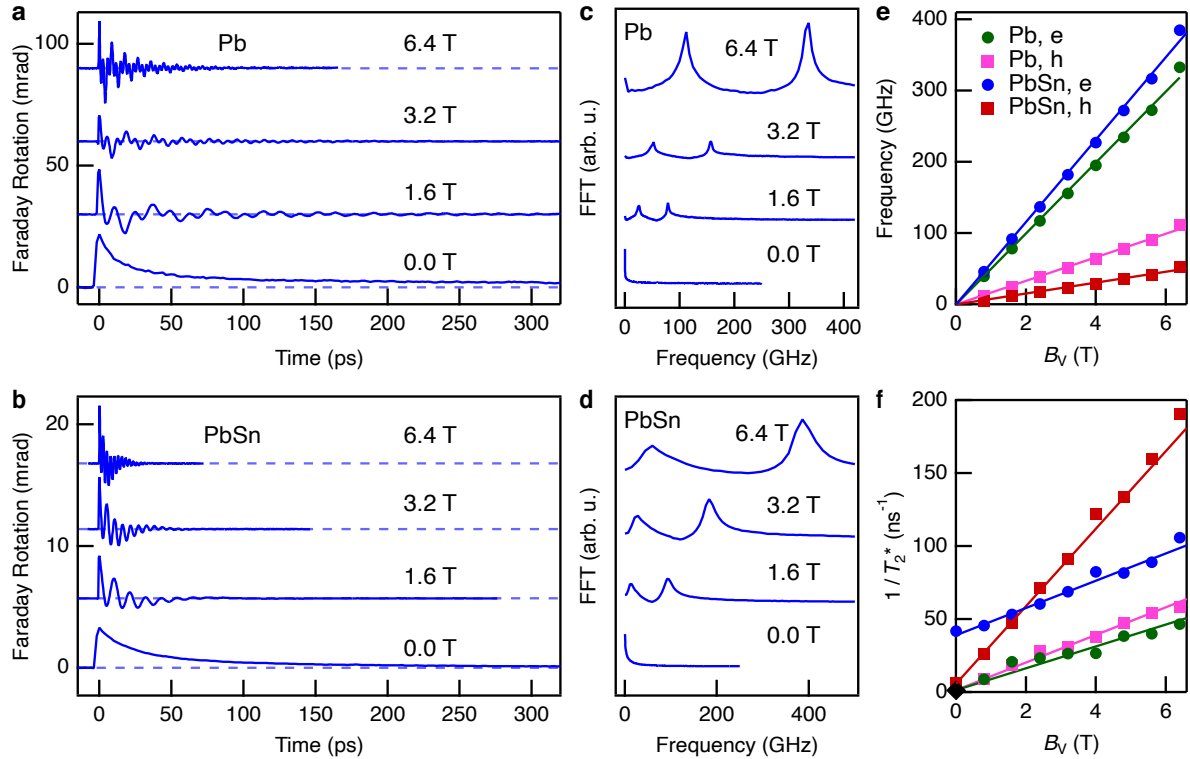

**Figure 2 a-d** Time-resolved Faraday rotation traces and their fast Fourier transform (FFT) spectra under different in-plane magnetic fields for the Pb- and PbSn-perovskites at 4 K, where arb. u. stands for arbitrary units. Curves are offset for clarity. Larmor precession frequency (**e**) and spin dephasing rate $1/T^*_2$ (**f**) versus the magnetic field $B_V$ for electrons and holes in the Pb- and PbSn-perovskites, extracted from the time-resolved Faraday rotation data (**a** and **b**). The black marker in panel **f** is the longer lifetime extracted from the double-exponential fit of the time trace for the Pb-perovskite at zero field. Error bars are smaller than the marker sizes in panels **e** and **f**.

The two oscillation frequencies represent the Larmor precession frequencies of electrons and holes: $f_{e,h} = |g_{e,h}|\mu_B B_V/h$, where $g_{e,h}$ is the electron (hole) g-factor, $\mu_B$ is the Bohr magneton, $B_V$ is the magnetic field strength and $h$ is the Planck constant. From the data in

Figure 2e, we obtain the absolute values of the g-factors. We further assign the g-factors and determine their signs (summarized in Table 1) according to the $\boldsymbol{k} \cdot \boldsymbol{p}$ model[10,11] (see below). In addition, the signs of the g-factors can be obtained from the exciton Zeeman splitting measurement[4,5,7,19] (see Supplementary Note 4). Since the Pb-perovskite has an exciton g-factor of $g_X = +2.16(0.11)$ (Supplementary Note 4) and $g_X = g_e + g_h$ (Ref.[5]), we get $g_e > 0$ and $g_h < 0$. Notably, the linear fits of the oscillation frequencies in Figure 2e have zero intercepts with the frequency axis, indicating a negligible exciton exchange field on electrons and holes[4].

**Table 1** g-factors $g$, g-factor spread $\Delta g$, zero-field spin lifetime $\tau_S$ for electrons and holes in the Pb- and PbSn-perovskites extracted from time-resolved Faraday rotation measurements at 4 K. The errors from the fits for both $g$ and $\Delta g$ are smaller than 0.001 and thus not shown. The error for $\tau_S$ is shown in the parentheses following the data.

| Material | | $g$ | $\Delta g$ | $\tau_S$ (ps) |
|---|---|---|---|---|
| Pb | e | 3.75 | 0.08 | 671 (33) |
| | h | -1.24 | 0.11 | 751 (42) |
| PbSn | e | 4.36 | 0.11 | 25.0 (0.1) |
| | h | -0.57 | 0.30 | 171 (2) |

The g-factor describes the Zeeman splitting between different spin states in the band edges and is thus determined by the band structure parameters. For the cubic-phase perovskites, a three-level $\boldsymbol{k} \cdot \boldsymbol{p}$ model[10,11] was developed, giving the analytical expressions $g_e = -\frac{2}{3} + \frac{4}{3}\frac{p^2}{m_0 E_g} + \Delta_{\text{rvb}}$ and $g_h = 2 - \frac{4}{3}\frac{p^2}{m_0}\frac{\Delta}{E_g(E_g + \Delta)}$, where $E_g$ is the bandgap, $\Delta$ is the spin-orbit splitting in the conduction bands, $p$ is the interband matrix element of the momentum operator, $m_0$ is the free-electron mass and $\Delta_{\text{rvb}}$ is the contribution to the electron g-factor from the remote valence bands.

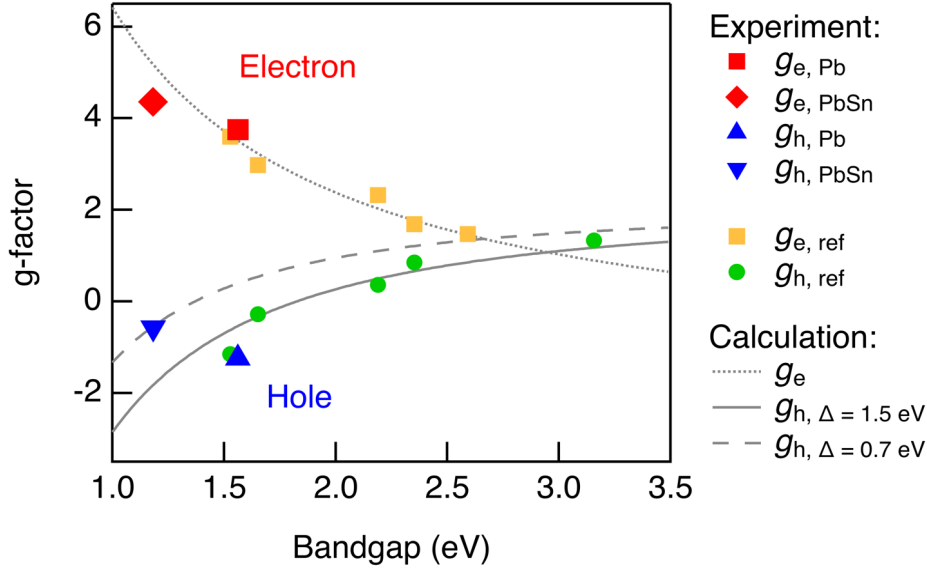

**Figure 3 Bandgap dependence of the electron and hole g-factors for the perovskites.** The red and blue markers are the g-factor values for our samples ($g_{e,Pb}$, $g_{h,Pb}$, $g_{e,PbSn}$ and $g_{h,PbSn}$) and the yellow and green markers are the g-factors for various lead-based perovskites with different compositions measured by Kirstein *et al.* in Ref.[11] ($g_{e,ref}$ and $g_{h,ref}$). The $g_{zz}$ values in Ref.[11] are shown here and the g-factor anisotropy is omitted. The curves are theoretical predictions from the $\bm{k} \cdot \bm{p}$ expressions in the main text. Δ denotes the spin-orbit splitting energy for the perovskites in the conduction band, whose value is Δ = 1.5 eV for the lead perovskites and Δ = 0.7 eV for the PbSn-perovskite from the best fit.

For the g-factors in lead-based perovskites, a universal dependence on the bandgap has been revealed based on these expressions[11]. By plotting our Pb-perovskite g-factors together with those from Ref.[11] in Figure 3, we find that they follow the same dependence on the bandgap. The electron g-factor of the PbSn-perovskite also obeys the same trend as the lead perovskites, even with its much smaller bandgap and SOC gap. However, the hole g-factor $g_{h,PbSn}$ deviates from the $g_h$ fit for the lead perovskites, highlighting that the hole g-factor depends sensitively on the SOC splitting energy Δ, consistent with the $\bm{k} \cdot \bm{p}$ model. We find that a reasonable value[26,27] of Δ = 0.7 eV fits $g_{h,PbSn}$, in comparison with Δ = 1.5 eV for the lead perovskites.

Note that the g-factors in these perovskites are generally larger than conventional semiconductors such as GaAs, ZnSe and CdTe with similar bandgaps[28–32]. The $g_e$ values in our samples are the largest among the reported g-factors in inorganic or hybrid perovskites, demonstrating that electron spins in these perovskites can be manipulated with smaller magnetic fields and are thus promising for spintronic applications.

Figure 2f shows that the spin decay rate increases linearly with the field for both electrons and holes in both samples, indicating that the decay is dominated by ensemble spin dephasing because of the inhomogeneous broadening of g-factors. The phenomena can be described by $1/T^*_{2,e,h} = 1/\tau_{S,e,h} + \Delta g_{e,h}\mu_B B_V/\hbar$ (Refs.[4,5,33]), where $\hbar$ is the reduced Planck constant, $\tau_{S,e,h}$ is the spin lifetime at zero field and $\Delta g_{e,h}$ is the spread of g-factors. The fits in Figure 2f using this expression give the g-factor broadening parameter $\Delta g$ and spin lifetime $\tau_S$, which are summarized in Table 1.

The g-factor spread $\Delta g$ is likely affected by the sample quality factors such as impurity level and spatial inhomogeneity. The $\Delta g_{e,h}$ of the Pb-perovskite are similar to those reported in Refs.[4,5] However, the PbSn-perovskite has larger $\Delta g$ values than the Pb-perovskite for both electrons and holes, especially a $\Delta g_h$ value as large as 0.30, which possibly links with the sample's poorer morphology and more defects due to the easy oxidation of $Sn^{2+}$ to $Sn^{4+}$ and Sn vacancies.[16]

The contrasting spin lifetimes $\tau_S$ between the two perovskites suggest that crystal defects play an important role in determining spin relaxation. Although the Pb-perovskite has larger SOC and thus more efficient spin relaxation is expected, the spin lifetimes for both electrons and holes in the Pb-perovskite are significantly longer compared to those in the PbSn-perovskite at 4 K. Given that the PbSn-perovskite has a much higher defect level[15], we speculate that this is because at low temperatures, the impurity scattering via the Elliott-Yafet mechanism is dominant and leads to much faster spin relaxation in the PbSn system. Moreover, in the Pb-perovskite, the spin lifetime of electrons is comparable to that of holes, in stark contrast to the PbSn-perovskite, where the spin lifetime of holes is more than six times longer than that of electrons. We suspect that this may originate from the hole doping in the PbSn sample due to Sn vacancies[15,34], leading to reduced scattering of holes with defects (due to screening) than that of electrons.

To study the temperature dependence of the spin lifetime, we fit the time-resolved Faraday rotation data with double-exponentials at elevated temperatures at zero magnetic field (see Supplementary Figure 2). The short-lived component for the Pb-perovskite, lasting ~30 ps at low temperatures, has weak temperature dependence (see Supplementary Figure 3) and does not exist in an applied transverse magnetic field, similar to the previous studies[4,5]. The longer lifetime (~800 ps at 4 K) drops significantly with increasing temperature and is related to the spin lifetimes of both electrons and holes because their dynamics are similar (see Table 1) and indistinguishable from the fits at zero field. For the PbSn-perovskite, by contrast, both the short and long lifetimes extrapolate well with the high-field $T_2^*$ data (see the zero-field data points in Figure 2f), so we identify them as the electron and hole spin lifetimes.

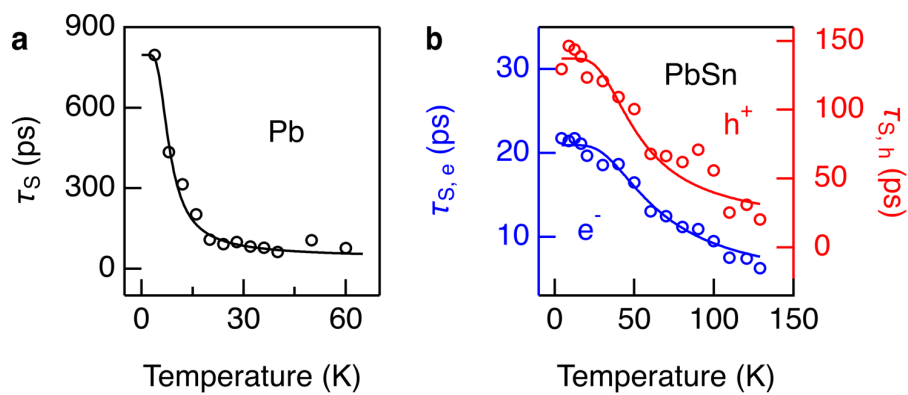

**Figure 4** Temperature dependence of the spin lifetime at zero magnetic field for the Pb- and PbSn-perovskites. The solid lines are Arrhenius activation-law fits to the data. The fits yield the activation energy of 2.3 meV for the Pb-perovskite and 12.2 meV for the PbSn-perovskite. Error bars are smaller than the marker size.

The spin lifetimes show strong temperature dependence and diminish with increasing temperature for both samples, as shown in Figure 4. Similar results have been observed in other perovskite systems[4–7,13] and were described by the Arrhenius activation-law expression $\frac{1}{\tau_S(T)} = \frac{1}{\tau_S(T=0)} + w\exp(-\frac{\Delta E}{k_B T})$, where $\tau_S(T=0)$ is the zero-temperature spin lifetime, $w$ is a constant and $\Delta E$ is the activation energy. The expression typically represents that spin scattering in the system has a clearly-defined energy barrier that needs to be thermally activated. With the same model, $\Delta E = 2.3$ meV is extracted for the Pb-perovskite and $\Delta E = 12$ meV is extracted for both electrons and holes in the PbSn-perovskite. This energy scale may be related to thermal excitation out of trapped states[7,35] or due to phonon-induced spin relaxation mechanisms[4–6,8,13]. On the one hand, trap states are easily formed in these solution-processed perovskite samples, especially the Pb and Sn vacancies[34,35]. On the other hand, transport measurements have indicated that momentum scattering is dominated by acoustic phonon scattering[36] and the longitudinal optical phonon modes have been measured to be ~10 meV (Refs.[37,38]) in lead perovskites. Future experimental and theoretical work considering the perovskite's unique band structures and carrier scattering with defects and phonons is needed to clarify these effects.

In Figure 5a and b, we further measure the temperature dependence of the g-factors for both perovskites based on the spin precession frequencies under a 0.8 T transverse magnetic field (time traces in Supplementary Figure 2). The g-factors vary dramatically from 4 to 70 K: the electron g-factors decrease by 0.44 and 0.35, while the hole g-factors increase by 0.51 and 0.13 in the Pb- and PbSn-perovskites, respectively. In stark contrast, the electron g-factor only changes by ~0.04 in GaAs and CdTe[39–42] and ~0.01 in InP[42] and GaN[43] over the same temperature range.

To determine the origin of the large g-factor shifts, we first examine the effect from the bandgap change with temperature. Extracted from absorbance spectra, the bandgap for both perovskites changes approximately linearly with the temperature from 10 to 70 K (Ref.[20] and Supplementary Figure 4). If temperature-independent values for the SOC gap and the interband matrix element are assumed, the bandgap change alone is estimated to result in a variation of -0.05 in $g_\mathrm{e}$ and 0.05 in $g_\mathrm{h}$ for the Pb-perovskite and -0.18 in $g_\mathrm{e}$ and 0.12 in $g_\mathrm{h}$ for the PbSn-perovskite, which are insufficient to describe the observed giant g-factor shifts, especially for the Pb-perovskite (see Supplementary Note 6 for details).

The strong temperature dependence of the g-factors is likely a result of the combined modification not only in the bandgap, but also in the SOC gap and the interband transition matrix elements. By inducing lattice deformations, thermal effects can cause significant changes in the perovskite's band structure. Well-studied mechanisms include the lattice thermal expansion[44], octahedral tilting of the inorganic perovskite framework[20,45,46], anharmonic lattice vibrations[20,47–49], etc.

To elucidate these structural effects on the g-factors, we build an empirical $sp^3$ tight-binding model based on the existing parameters for the cubic-phase MAPbI$_3$ (Ref.[24]), which has very similar bandgap and SOC gap values as our Pb-perovskite. Details are included in Supplementary Note 6. We find that the dominant contribution results from the lattice vibration effects on the band structure, whereas the lattice expansion contributes a very small portion of the observed g-factor shift (less than 1/5, see Supplementary Figure 5).

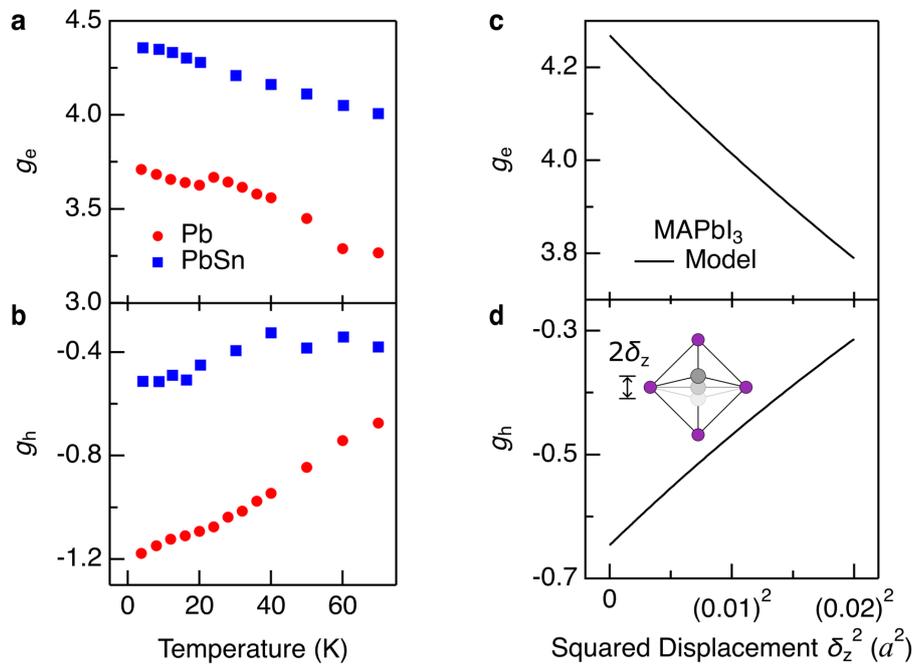

**Figure 5 a** and **b** Temperature dependence of the electron and hole g-factors, respectively, extracted from the time-resolved Faraday rotation data at 0.8 T (Voigt geometry). **c** and **d** The electron and hole g-factors as a function of the square of the vibration amplitude $\delta_z^2$ for the Pb-atoms in cubic MAPbI$_3$, calculated by an empirical tight-binding model and the g-factor expressions in Ref.[11] (details in Supplementary Note 6), where $a$ is the lattice constant. The inset in panel **d** shows a schematic of the lattice vibration, with the gray ball being the lead atom and the purple balls being the iodine atoms.

We calculate the impact on the band structure and the g-factors from the possible thermal vibration modes represented by the dynamic displacements of the lead and iodine atoms. An off-centering shift of the lead atom in the $PbI_6^-$ octahedra is depicted in the inset of Figure 5d and other modes are shown in Supplementary Figure 6. Characterized by the square of the atomic displacement $\delta_z^2$, the growing thermal vibrations cause a substantial increase in the bandgap while decreasing both the SOC gap and the interband transition matrix elements as shown in Supplementary Figure 6 (g-i). Overall, in Figure 5c and d, when the vibration amplitude reaches as large as $0.02a$ ($a$ is the lattice constant), $g_e$ decreases by ~0.48 and $g_h$ increases by ~0.33, comparable to our experiment results. The displacement range up to $0.02a$ is reasonably chosen based on the ab initio molecular dynamics simulation in Ref.[47] of the mean square displacement for the Pb atoms in $CsPbBr_3$ at 150 K, which is about 0.033 Å$^2$. The value can be linearly extrapolated to 0.015 Å$^2$ at 70 K, corresponding to $\sim(0.02a)^2$ in our Pb-perovskite. At low temperatures, the thermal energy is proportional to the square of the lattice vibration amplitude[47], which in turn induces almost linear shifts of g-factors with the temperature, in excellent agreement with the experimental results.

In summary, we have done a comparative study of spin coherence, spin dephasing and spin relaxation of electrons and holes in the Pb- and PbSn-perovskites using time-resolved Faraday rotation spectroscopy. From these measurements we have not only determined the Landé g-factors and spin lifetimes but also shed light on their thermal evolutions. While the electron and hole g-factors for the Pb-perovskite and the electron g-factor for the PbSn-perovskite follow the theoretical predictions developed for lead perovskites, a modified SOC energy in the model is required to fit the hole g-factor for the PbSn-perovskite. The relatively small bandgap in our samples makes their electron g-factors the largest among the reported values in inorganic or hybrid perovskites. While a long-lived subnanosecond spin relaxation and spin coherence of electrons and holes have been observed in the Pb sample, the spin lifetimes in the PbSn-perovskite suffer significantly from its higher defect level despite its smaller SOC. In addition, we have pointed out that the strong temperature dependence of the g-factors likely originates from the lattice thermal vibrations, which greatly modify the bandgap, the SOC gap and the momentum operator transition matrix elements. Our findings lay the foundation for future spintronic applications based on the tin perovskites and provide insights into the unique thermal effects on the spin coherence and g-factors in the perovskite systems.

## Supporting Information

Supplementary Notes on Sample Preparation, Absorption Measurements, Time-resolved Faraday Rotation Measurements, Exciton Zeeman Splitting, Temperature Dependence of Time-resolved Faraday Rotation Data, Thermal Shifts of g-factors in the Perovskites Induced by Lattice Distortions

## Acknowledgements

Samples were prepared at Nanjing University. All optical measurements and calculations were performed at Tsinghua University. L.Y. acknowledges the support from the National Key R&D Program of China (Grant Nos. 2020YFA0308800 and 2021YFA1400100) and the National Natural Science Foundation of China (Grant Nos. 12074212). Y.X. was supported by the National Key R&D Program of China (2018YFA0307100 and 2018YFA0305603) and the National Natural Science Foundation of China (12025405 and 11874035).  The work of H.T. was supported by the National Natural Science Foundation of China (Grant Nos. 61974063 and U21A2076). H.Z. was also supported by funds from the University of Toronto.

## Competing interests

The authors declare no competing interests.

**Supporting Information**

**Supplementary Note 1: Sample Preparation**

The samples are solution-processed $MA_{0.3}FA_{0.7}PbI_3$ and $MA_{0.3}FA_{0.7}Pb_{0.5}Sn_{0.5}I_3$ perovskite thin films. Both samples keep a cubic perovskite structure from 90 to 340 K based on our X-ray diffraction, photoluminescence and absorbance measurements[1]. The thicknesses of the Pb- and PbSn-perovskites are 1.0 and 1.2 µm respectively. The typical grain size for both perovskite samples is 1 µm. The PbSn-perovskite was prepared with a tin-reduced precursor solution strategy to effectively prevent the oxidation of $Sn^{2+}$. Characterized in Ref.[2], the synthesis strategy of the PbSn-perovskite significantly reduces both the trap density and the hole density to half of the value compared to the control sample. The PbSn-perovskite has achieved a record power conversion efficiency when implemented in an all-perovskite tandem solar-cell[2].

## Supplementary Note 2: Absorption Measurements

The absorbance spectrum was measured in a closed-cycle cryostat equipped with a superconducting magnet. The white light from a stabilized quartz tungsten-halogen broadband light source was directed to and focused on the sample using free space optics. The circularly polarized white light was achieved by letting the beam pass a broadband polarizer and a quarter-waveplate whose fast axis is 45° rotated from the polarizer axis. The light beam was aligned with the magnetic field direction so that the differences between the light absorption with opposite helicities caused by the Zeeman splitting can be measured. The transmitted light was collected with an optical fiber coupled to a spectrometer. We used a 1200 grooves/mm blazed at 500 nm grating to disperse the light and analyzed the spectrum with an EMCCD camera (for the measurements of the Pb-perovskite). The system has a resolution of 0.02 nm. The absorbance spectrum of the PbSn-perovskite sample was measured with the same experimental setup except with a single-point InGaAs detector. The EMCCD camera provides exceptional sensitivity and noise performance (~single electron) compared to the InGaAs detector (~$10^3$ electrons), which makes the Zeeman splitting measurement on the Pb-perovskite possible. We also measured the temperature dependence of the absorbance for both perovskite samples to trace their bandgap shift with temperature.

## Supplementary Note 3: Time-Resolved Faraday Rotation Measurements

The time-resolved Faraday rotation experiment was conducted with a wavelength-tunable 80 MHz femtosecond Ti:Sapphire oscillator. The laser beam was split into the pump and probe beam paths. The pump beam was modulated between right- and left-circularly polarized states by a photo-elastic modulator to facilitate lock-in measurements. The time-resolved Faraday rotation experiment was set up around the same cryostat as in the absorbance measurements but in the Voigt geometry, where the laser beam propagation direction is normal to the sample plane and the magnetic field is in the sample plane. The laser was tuned to the wavelength near the bandgap (~817 nm for the Pb-perovskite, ~1020 nm for the PbSn-perovskite). Because the bandgap of the perovskites is sensitive to temperature, we tuned the laser wavelength accordingly based on our absorbance measurements to trace the bandgap shift. Both the pump and probe beams were focused on the sample using the same lens with a ~80 μm spot (pump fluence ~0.2 μJ cm$^{-2}$).

## Supplementary Note 4: Exciton Zeeman Splitting

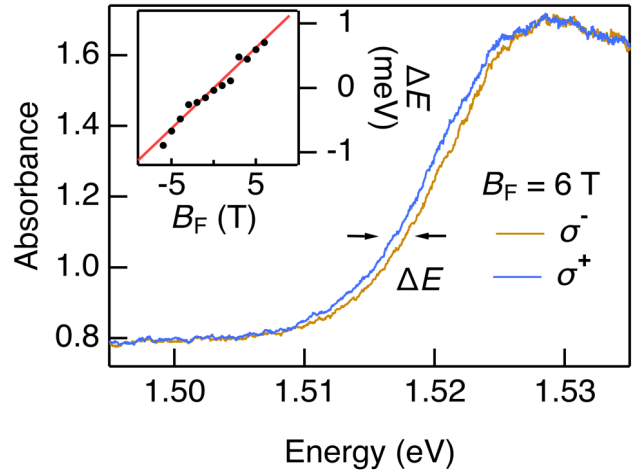

**Supplementary Figure 1 Exciton Zeeman splitting for the Pb-perovskite.** Absorbance spectra with opposite helicities of circularly polarized light ($\sigma^-$ and $\sigma^+$) under a longitudinal magnetic field $B_\mathrm{F} = 6$ T (Faraday geometry) for the Pb-perovskite at 4 K. The energy splitting between $\sigma^-$ and $\sigma^+$ absorbance spectra denoted by $\Delta E$ is the Zeeman splitting of the excitons. The inset shows the linear relationship between $\Delta E$ and $B_\mathrm{F}$, whose slope gives the exciton g-factor.

Under a longitudinal magnetic field $B_\mathrm{F}$ (the subscript F stands for the Faraday geometry, i.e., the field is aligned with the light beam propagation direction), the energies of the spin up and down states move in the opposite directions, resulting in a difference in the absorption between left- and right-circularly polarized light near the band edge. This difference is attributed to the exciton Zeeman splitting, which is shown in Supplementary Figure 1 by plotting the absorption onset of light with opposite helicities at $B_\mathrm{F}$ = 6 T in the Pb-perovskite. The splitting energy $\Delta E$ changes linearly with the magnetic field $B_\mathrm{F}$ (inset), corresponding to the expression $\Delta E = g_\mathrm{X} \mu_\mathrm{B} B_\mathrm{F}$, where $g_\mathrm{X}$ is the exciton $g$-factor and $\mu_\mathrm{B}$ is the Bohr magneton. We extract the exciton g-factor $g_\mathrm{X} = 2.16(0.11)$ from the slope and confirm the positive sign of $g_\mathrm{X}$ by explicitly comparing the direction of the magnetic field with the helicity of the light. This $g_\mathrm{X}$ value confirms the recent finding of a weak dispersion of the exciton g-factor with the bandgap energy in lead perovskites[3].

We do not observe an obvious Zeeman splitting in the PbSn-perovskite due to the much broader absorption edge (Figure 1b in the main text) and the much lower signal-to-noise ratio of the InGaAs detector used in the PbSn-perovskite absorption measurements compared with the case of the Pb-perovskite (see Supplementary Note 2).

# Supplementary Note 5: Temperature Dependence of Time-Resolved Faraday Rotation Data

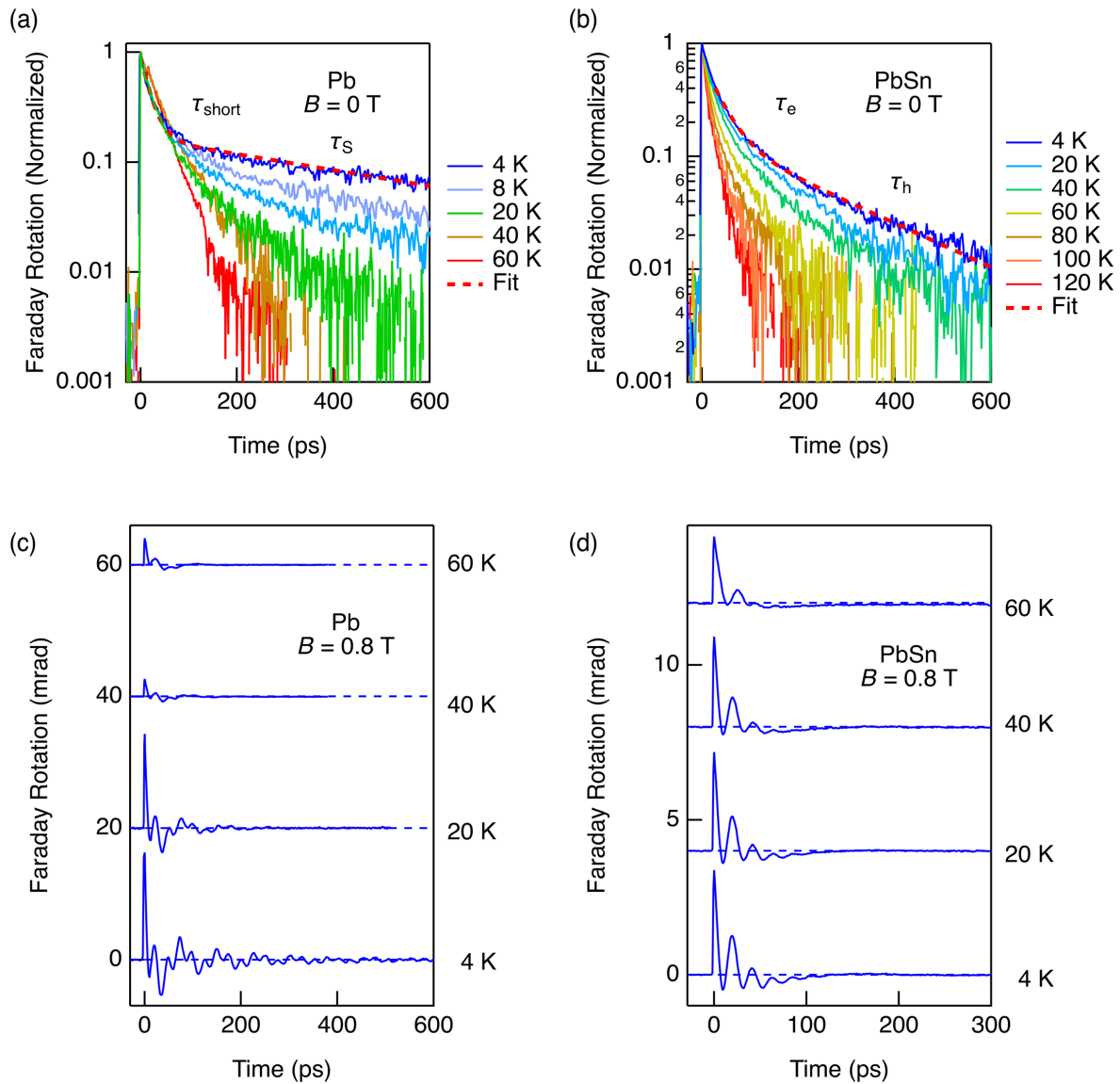

**Supplementary Figure 2 Temperature dependence of the time-resolved Faraday rotation data at zero field (a,b) and 0.8 T (c,d).** The time traces in (a) and (b) are fitted with double-exponentials. For the Pb-perovskite, the longer lifetime corresponds to the spin lifetime $\tau_S$, where electron and hole spin lifetimes are indistinguishable at zero magnetic field; for the PbSn-perovskite, two lifetimes are identified as the electron and hole spin relaxation times, $\tau_e$ and $\tau_h$, respectively.

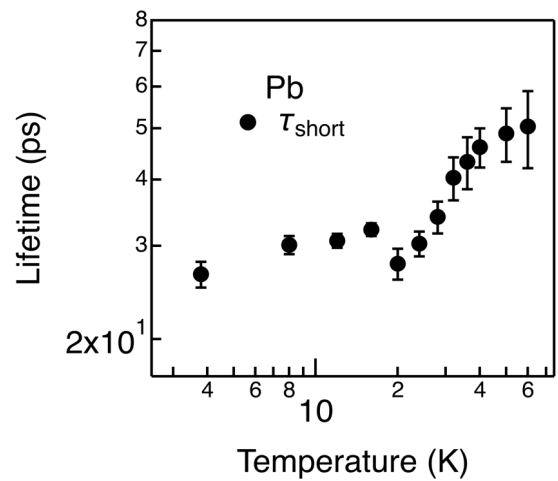

**Supplementary Figure 3** Temperature dependence of the short lifetime $\tau_{short}$ for the Pb-perovskite.

# Supplementary Note 6: Thermal Shifts of g-Factors in the Perovskites Induced by Lattice Distortions

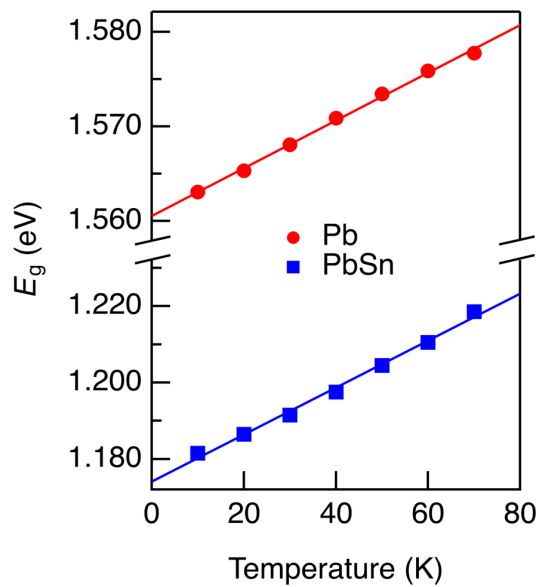

**Supplementary Figure 4 Temperature dependence of the bandgap in the Pb- and PbSn-perovskites.** The bandgap values are extracted from the absorbance curves at the corresponding temperatures using the Elliot model[1]. The solid lines are linear fits for the temperature dependence of the bandgap for both perovskites.

The analytical expressions in the main text for the g-factors show how they are connected to the bandgap, the SOC gap and the interband transition matrix elements.

As shown in Supplementary Figure 4, the measured bandgap of the Pb- and the PbSn-perovskites increases with temperature from 10 to 70 K with the rates of roughly 0.25 and 0.61 meV K$^{-1}$, respectively. However, the bandgap shifts alone are insufficient to account for the huge g-factor changes in this temperature range. For example, in the Pb-perovskite, the bandgap needs to shift at least eight times faster with temperature to achieve the measured change in their g-factors while keeping all the other parameters fixed.

We propose that not only the bandgap but also the SOC gap and the interband transition matrix elements vary significantly with the temperature, and the thermal lattice distortions are the main source causing the band structure shift. To support our claims, we build an empirical $sp^3$ tight-binding model based on the parameters given in Ref.[4] for cubic MAPbI$_3$ and the matrix elements of the momentum operator are calculated with a standard approach[5,6]. We consider the influence of the lattice distortions on the band structure and the interband transition matrix.

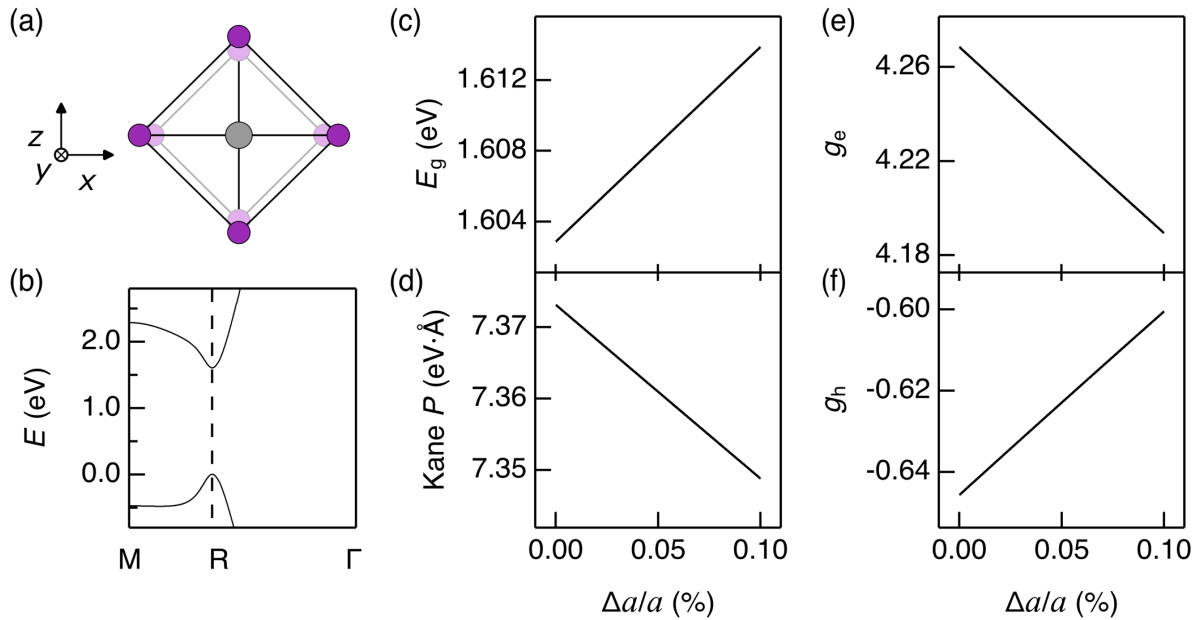

**Supplementary Figure 5 Lattice dilation effects on the bandgap and the g-factors of MAPbI$_3$ by the tight-binding model.** (a) Schematic of the lattice dilation of the PbI$_6^-$ octahedron. (b) Band structure under the lattice dilation corresponding to (a). (c-f) Calculation results for the bandgap $E_g$, the interband transition matrix element $p$ (the Kane matrix element $P = \hbar p/m_0$) and the g-factors $g_e$ and $g_h$ due to the lattice constant expansion up to $\frac{\Delta a}{a} = 0.1\%$.

First, we examine the lattice dilation effects. The thermal lattice expansion coefficient is measured to be 1.40*10$^{-5}$ and 0.72*10$^{-5}$ K$^{-1}$ for Pb and PbSn samples, respectively, from X-ray diffraction measurements, resulting in a lattice expansion of 0.08% and 0.04% from 10 to 70 K. In the tight-binding calculations, the lattice expansion reduces the transfer matrix elements between different atoms, which are inversely proportional to the square of the lattice constant[7,8]. Supplementary Figure 5 demonstrates the calculated results for thermal lattice expansion up to ~0.1%. The bandgap changes less than half of the experiment results, and, especially, the shifts in $g_e$ and $g_h$ are much smaller than the experimental values in the Pb-perovskite (Figure 5 in the main text). Note that the SOC gap (not shown) changes marginally in this lattice expansion range. Therefore, we instead consider the lattice vibration effects while retaining the equilibrium cubic perovskite structure.

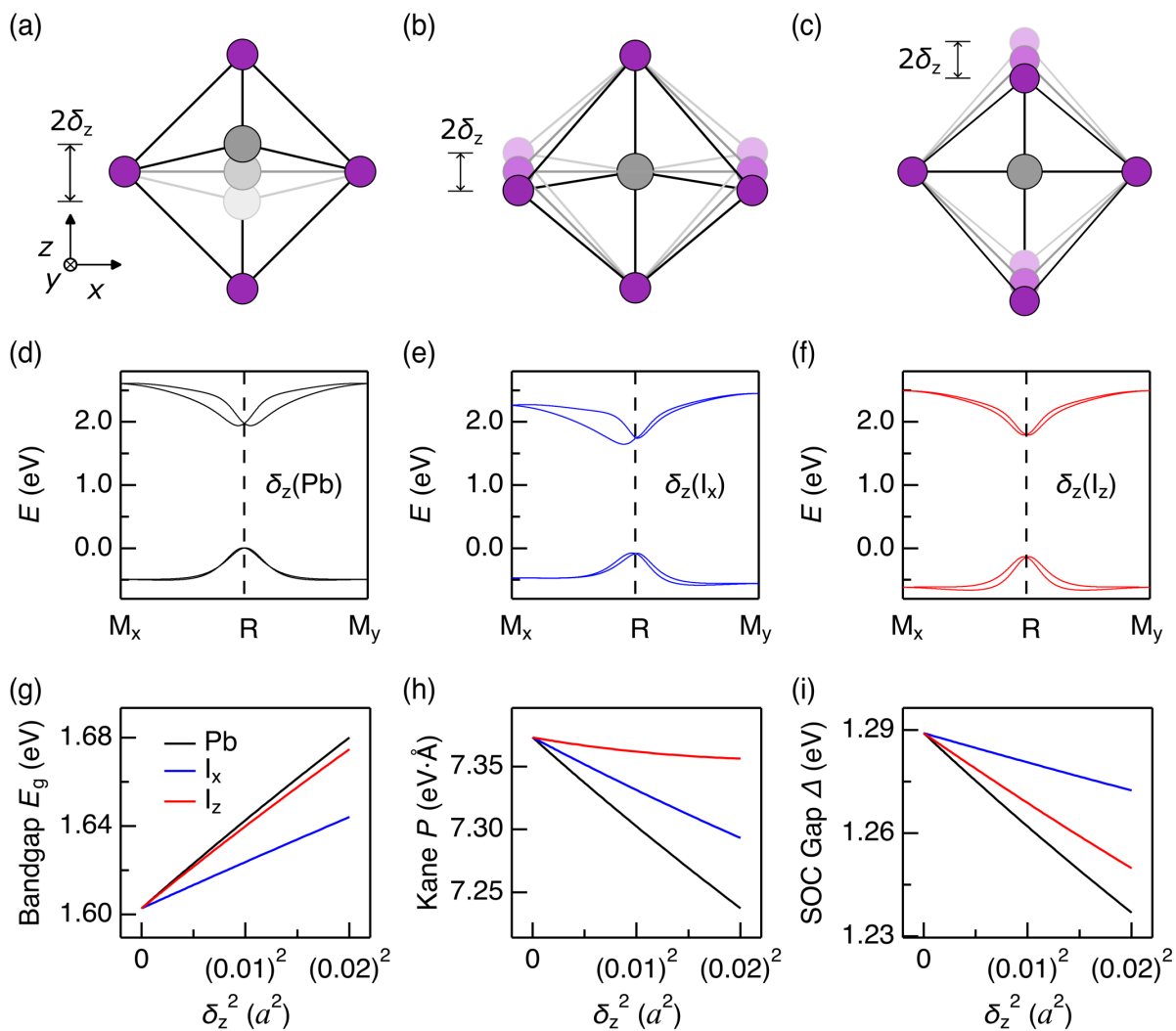

**Supplementary Figure 6 Lattice vibration effects on the band structure of MAPbI$_3$ by the tight-binding model.** (a-c) Schematics of the three vibration modes we consider in the calculation, where $\delta_z$ denotes the displacement of the corresponding atoms along the z direction. The three modes are the vibration of the lead-atom Pb (a), the iodine atoms on the x-axis I$_x$ (b) and the iodine atoms on the z-axis I$_z$ (c). (d-f) Band structures along the M$_x$-R-M$_y$ direction under the vibration modes corresponding to (a-c). (g-i) Vibration-induced shifts in the bandgap $E_g$, the interband transition matrix element $p$ (the Kane matrix element $P = \hbar p/m_0$) and the SOC gap $\Delta$. The direction-averaged Kane parameter ($P = \hbar/m_0\sqrt{(p_x^2 + p_y^2 + p_z^2)/3}$) is calculated considering that we measure polycrystalline samples.

The perovskites have relatively "soft" lattice structures susceptible to dynamic vibration effects[9,10], which cause strong modifications of their band structure. In the cubic-phase perovskites, this dynamic effect of the local symmetry breaking has been shown to be originated from the anharmonicity in the potential energy surface and is associated with the soft phonon modes at high temperatures[11,12]. Previous molecular dynamics simulations on inorganic perovskites[10–12] showed that the thermal vibrations could distort the lattice bond angle by more than 10° at room temperature, and the bandgap renormalization due to the electron-phonon coupling was substantial to cause bandgap shifts by hundreds of meV. Because the state-of-the-art treatments of the anharmonic effects on the band structure require sophisticated first-principles calculations[13], we instead provide a simple picture of the lattice vibration effects on the g-factors based on the tight-binding model.

We examine three representative vibration modes by displacing the lead and two inequivalent iodine atoms (in the x and z directions relative to the Pb atom) in the $PbI_6^-$ octahedra along the z-direction and schematically display them in Supplementary Figure 6(a-c). Supplementary Figure 6(d-f) show their corresponding band structures near the R point, where the bandgap is located. Due to the broken inversion symmetry induced by these distortion modes, the Rashba splitting occurs near the band edges (except along the R-$M_z$ direction). Note that the four-fold rotational symmetry about the z-axis is retained in the cases shown in (a,d;c,f), but broken in the case shown in (b,e).

In all three modes, when the vibration grows, the bandgap increases while the SOC gap and the transition matrix element decrease as shown in Supplementary Figure 6(g-i), all pointing towards the measured shifting trends of $g_e$ and $g_h$ considering the expressions of the g-factors in the main text. As a result, all three modes cause a decrease of $g_e$ and an increase of $g_h$ (Supplementary Figure 7). By choosing a range up to $0.02a$ for $\delta_z$, which is the estimated vibration amplitude for lead atoms at 70 K (reason stated in the main text), the mode with lead displacements induces shift of about -0.48 in $g_e$ and 0.33 in $g_h$, in good agreement with the experimental results. Although the vibrations in iodine atoms induce smaller shifts in g-factors, their larger vibration amplitudes due to the smaller atomic mass can compensate this and cause similar g-factor shifts as the case considering the lead atom vibrations.

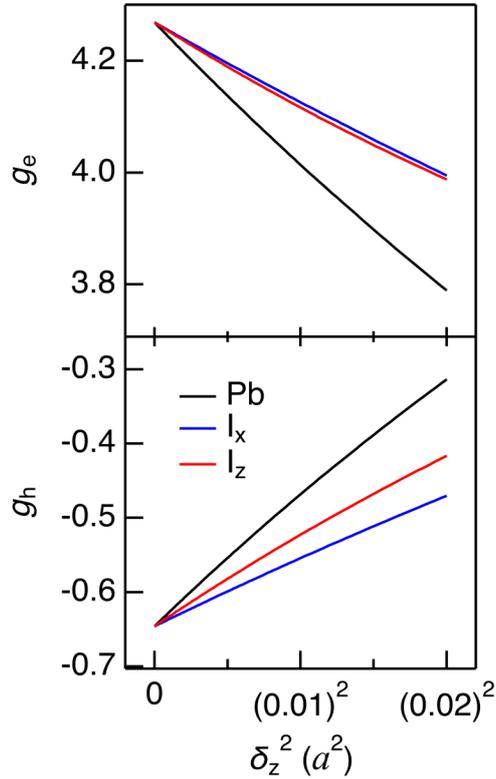

**Supplementary Figure 7 Lattice vibration effects on the g-factors of MAPbI$_3$ by a tight-binding model.** The three vibration modes are shown schematically in Supplementary Figure 6 (a-c).

Note that in GaAs the mechanism of the electron g-factor shift with temperature is still under debate. The experimental observations of the g-factors contradicted the initial prediction of the $\mathbf{k} \cdot \mathbf{p}$ theory if only the bandgap change with temperature was considered[14]. It turned out that the contributions from the temperature dependence of the interband matrix due to dynamic lattice and the remote bands and non-parabolicity of the conduction band are also important[15,16]. A 5.4% decrease of the Kane energy was required to fit the experimental g-factor from 2.6 K to room temperature[15].

**Supplementary References**